\documentstyle[12pt]{article}
\begin{document}
\parskip 10pt plus 1pt
\title{Quantum Logic Gates using q-deformed Oscillators}
\author{
{\it Debashis Gangopadhyay}\\
{S.N.Bose National Centre For Basic Sciences}\\
{JD Block, Sector-III, Salt Lake, Kolkata-700098, INDIA}\\
{\it debashis@boson.bose.res.in}\\
{\it Mahendra Nath Sinha Roy}\\
{Dept.of Physics, Presidency College}\\
{86/1, College Street,Kolkata-700073, INDIA}
}
\date{}
\maketitle
\baselineskip=20pt
\begin{abstract}
We show that the quantum logic gates ,{\it viz.} the 
single qubit Hadamard and Phase Shift gates, can also be realised 
using  q-deformed angular momentum states
constructed via the Jordan-Schwinger mechanism with two q-deformed oscillators. 


\end{abstract}
\newpage

{\bf 1. Introduction}

Quantum logic gates are basically unitary operators (Refs 1-4 and 
references therein).
There are two gates , the Hadamard and Phase Shift gates, 
which are sufficient to construct any unitary operation
on a single qubit$^{5-7}$.These gates are constructed
using the "spin up" and "spin down" states of
$SU(2)$ angular momentum i.e., the two possible states of a qubit are  usually 
represented by "spin up" and "spin down" states. In this work 
we show that the Hadamard and Phase Shift gates can also be realised
with {\it q-deformed angular momentum states
constructed via Jordan-Scwinger mechanism with two q-deformed oscillators}. 
We employ the technique of harmonic oscillator realisation of q-oscillators$^{12-17}$.

The motivation of our work comes from the fact that there exists a non-trivial
generalisation$^{12-13}$ of the harmonic oscillator realisation of q-oscillators. This
generalised scheme allows us to set up an alternate quantum computation 
formalism at the level of choosing the two basis states. Consequently,
this formalism is more general and contains the currently used  formalism in quantum 
computation as a special case, i.e. for $q=1$.Let us clarify this further.
$a_{q}^{\dagger}$ and $a_{q}$ are the creation and annihilation operators for
q-oscillators while those for the usual oscillators are $a^{\dagger}$ and $a$.
These satisfy (with $q= e^{s}$, $0\leq s\leq 1$):
$$a_{q}a_{q}^{\dagger}- qa_{q}^{\dagger}a_{q}=q^{-N}~~ ;~~ N^{\dagger}=N\eqno(1a)$$
$$[N,a_{q}]=-a_{q}~;~ [N,a_{q}^{\dagger}]=a_{q}^{\dagger}~;~ a_{q}^{\dagger}a_{q} = [N];~
a_{q}a_{q}^{\dagger}=[N+1]\eqno(1b)$$
$$a_{q}f(N) = f(N+1) a_{q}~~;~~a_{q}^{\dagger}f(N)=f(N-1)a_{q}^{\dagger}\eqno(1c)$$
where $[x]=(q^{x}-q^{-x})/(q-q^{-1})$ and $N$ is the number operator (eigenvalue $n$)
for the q-deformed oscillators and $f(N)$ is any function of $N$.The above equations 
are true for both real and 
complex $q$. However, we shall confine ourselves to real $q$ $^{10,11}$.
The harmonic oscillator realisation of quantum oscillators$^{12,13}$ gives the 
relationships between
$a_{q},a_{q}^{\dagger}$ and  $a,a^{\dagger}$ as
$$a_{q}=a\sqrt{{\frac {q^{\hat N} \psi_{1} - q^{-\hat N}\psi_{2}}{{\hat N (q-q^{-1})}}}}\enskip;\enskip
a_{q}^{\dagger}=\sqrt{{\frac {q^{\hat N} \psi_{1} - q^{-\hat N}\psi_{2}}{{\hat N (q-q^{-1})}}}} a^{\dagger}\eqno(2a)$$
$$ N = \hat N - (1/s) ln~ \psi_{2}\eqno(2b)$$
$\hat N$ is the number operator for usual oscillators with eigenvalue $\hat n$;
and $\psi_{1}$ , $\psi_{2}$ are arbitrary functions of $q$ only with $\psi_{1,2}(q)= 1$ for $q=1$.
{\it The presence of these arbitrary functions allows an alternative formalism:}

{\bf Case I :}
If all these arbitrary functions are unity, then $N=\hat N$.This means that
if states are labelled by their occupation numbers,deformed states cannot
be distinguished from the non-deformed (i.e. usual) oscillator states. This is
the realm of quantum computation with the the usual "spin-up" and "spin-down" 
states and there is no theoretical gain by choosing deformed oscillator states
as basis for quantum computation.  

{\bf Case II :}
However , the harmonic oscillator realisation    
$(2)$ is general if the arbitrary functions of $\psi_{i}(q), i=1,2$ are
{\it not all equal to unity}. Let us take $\psi_{1}=\psi_{2}=\psi(q)$ .Now 
$N=\hat N - (1/s)~ln~ \psi(q)$ (equation $(2b)$). 
Hence at the occupation number level states are different as the eigenvalues of the
number operator of  usual oscillator states (i.e. usual quantum computation) and the
eigenvalues of the number operator of deformed oscillator states are now related by
$n=\hat n - (1/s)~ ln ~\psi(q)$. This shows up in the Jordan-Schwinger construction
of angular momentum states and the states in the two cases will be distinguishable through
the function $\psi(q)$. So there is this  extra functional parameter $\psi(q)$
which is potentially ideal  for {\it experimental realisations}.

{\bf 2. Jordan-Schwinger construction for qubits}

We now discuss how qubits look in the Jordan-Schwinger construction where 
two independent oscillators are used  to construct the generators of angular momementum.

(a) States are  defined by the total angular momentum $j$ and
$z$-component of angular momentum  $j_{z}$,  
$$\vert j m>= {\frac {( a^{\dagger}_{1})^{j+m} (a^{\dagger}_{2})^{j-m}}{[(j+m)!(j-m)!]^{1/2}}}
\vert\phi> \eqno(3)$$
$\vert \phi> \equiv \vert\tilde 0> = \vert\tilde 0>_{1} \vert\tilde 0>_{2}$ is
the ground state ($j=0,m=0$). $\vert\tilde 0>_{i}, i=1,2$ are the oscillator
ground states.
$j=(n_{1}+n_{2})/2\enskip;\enskip m=(n_{1}-n_{2})/2$ and $n_{1},n_{2}$ are the 
eigenvalues of the number operators of the two oscillators.

(b) For qubits , the only possible states correspond to
$(n_{1}+n_{2})/2 = 1/2$ i.e. $n_{1}=1-n_{2}$.
States characterised by these are therefore 
$\vert (n_{1} + n_{2})/2 , (n_{1} - n_{2})/2 >\equiv \vert n_{1}> 
\vert n_{2}>\delta_{n_{1}+n_{2},1}$.
Since $j=1/2$ for both qubit states, we suppress $j$ and write the states as 
$$\vert  m>={\frac {(a^{\dagger}_{1})^{1/2 + m} (a^{\dagger}_{2})^{1/2  - m}}
{[(1/2  + m)!(1/2  - m)!]^{1/2}}}\vert\phi> \eqno(4a)$$
$$\vert - m>={\frac {(a^{\dagger}_{1})^{1/2 - m} (a^{\dagger}_{2})^{1/2  + m}}
{[(1/2  + m)!(1/2  - m)!]^{1/2}}}\vert\phi> \eqno(4b)$$
Equivalently, in terms of $n_{1},n_{2}$ these are 
$$\vert n_{1} -  1/2> ={\frac {(a^{\dagger}_{1})^{n_{1}} (a^{\dagger}_{2})^{1  - n_{1}}}
{[(n_{1})!(1 - n_{1})!]^{1/2}}}\vert\tilde 0> \eqno(4c)$$
$$\vert - (n_{1} -  1/2)> ={\frac {(a^{\dagger}_{1})^{1-n_{1}} (a^{\dagger}_{2})^{n_{1}}}
{[(n_{1})!(1 - n_{1})!]^{1/2}}}\vert\tilde 0> \eqno(4d)$$

(c)In this formalism  the two basis states of a single qubit state are
($\vert 1>\equiv \vert up>$ state and $\vert 0>\equiv \vert down>$ state )
$$\vert 1>\equiv\vert 1/2,1/2>
\equiv\vert 1/2>=a^{\dagger}_{1}\vert \tilde 0>
=a^{\dagger}_{1}\vert\tilde 0>_{1}\vert\tilde 0>_{2}
=\vert\tilde 1>_{1}\vert\tilde 0>_{2}$$
$$\vert 0>\equiv\vert 1/2,-1/2>
\equiv\vert -1/2>= a^{\dagger}_{2}\vert \tilde 0>
=a^{\dagger}_{2}\vert\tilde 0>_{1}\vert\tilde 0>_{2}
=\vert\tilde 0>_{1}\vert\tilde 1>_{2}$$

(d)The {\it physical meaning} of the notation is as follows. The $\vert 1>$
angular momentum (spin "up") state can be constructed out of two oscillator 
states where the first oscillator state has occupation number $1$ while the
other has occupation number $0$. The $\vert 0>$ ( spin "down") state
corresponds to the first oscillator having occupation number $0$ and the second 
oscillator having occupation number $1$. We thus can write any qubit state
in terms of harmonic oscillator states.
The column vectors denoting these two basis  states may be taken as 
$$\vert 1>=\pmatrix{1\cr 0\cr}\enskip ; \enskip 
\vert 0>=\pmatrix{0\cr 1\cr}$$
So we write
$$\vert x>=(a_{1}^{\dagger})^{x} (a_{2}^{\dagger})^{1-x}\vert \tilde 0>\eqno(5)$$
(Note $\vert 0>$ represents one of the two possible qubit states while 
$\vert\tilde 0>$ represents oscillator ground state i.e. occupation number $0$;$\vert\tilde 1>$ represents an oscillator state with occupation number $1$; $\vert\tilde 2>$
represents oscillator state with occupation number $2$ etc. This notation is to avoid confusion).

{\bf 3. The Hadamard transformation for q-deformed qubits}

First consider the case of an ordinary qubit.
The Hadamard transformation on a single qubit state ($x=0,1$) is$^{5-7}$
(modulo a normalisation factor of $ 1/\sqrt {2}$)
$$\vert x>\longrightarrow (-1)^{x}\vert x> ~+~\vert 1-x>\eqno(6)$$
Using $(4c), (4d), (5)$  in $(6)$ gives :
$$\vert n_{1} - 1/2>\longrightarrow (-1)^{n_{1}}\vert n_{1} - 1/2 > ~+~ \vert 1/2  -  n_{1}>\eqno(7)$$
So
$n_{1}=0\Rightarrow\vert -{\frac 12} >\longrightarrow \vert -{\frac 12}>~ +~ \vert {\frac 12}>$
and $n_{1}=1\Rightarrow \vert {\frac 12} >\longrightarrow \vert -{\frac 12}>~ -~ \vert{\frac 12}>$.

Now consider q-deformed qubits.
For states, we have kets $\vert >$ (or bras $<\vert$) for the usual oscillator states,
while  kets $\vert>_{q}$ (or bras $_{q}<\vert$) denote the corresponding q-deformed states.
The general angular momentum q-deformed state in terms two q-deformed
oscillators  is$^{8,9}$ 
$$\vert j m>_{q} \equiv {\frac {(a_{1q}^{\dagger})^{n_{1}} (a_{2q}^{\dagger})^{n_{2}}}{([n_{1}]![n_{2}]!)^{1/2}}} 
\vert\phi>_{q}\eqno(8a)$$
$$\vert j ~~~-m>_{q} \equiv {\frac {(a_{1q}^{\dagger})^{n_{2}} (a_{2q}^{\dagger})^{n_{1}}} 
 {([n_{1}]![n_{2}]!)^{1/2}}} \vert\phi>_{q}\eqno(8b)$$
where $\vert \phi>_{q}\equiv \vert\tilde 0>_{q}=\vert\tilde 0>_{1q}\vert\tilde 0>_{2q}$ 
is the ground state corresponding to two non-interacting
q-deformed oscillators. Ground states of q-oscillators
in the coordinate representation were studied in Refs. 8 and 9.
In our notation a qubit  state has either  (a) $n_{1}=0 , n_{2}=1$ or 
(b) $n_{1}=1 , n_{2}=0$. Thus from  $(8a),(8b)$ {\it the q-deformed
qubit would look like}
$$\vert n_{1} -  1/2 >_{q}
\equiv {\frac {(a_{1q}^{\dagger})^{n_{1}} (a_{2q}^{\dagger})^{1-n_{1}}} 
 {([n_{1}]![1-n_{1}]!)^{1/2}}} \vert \tilde 0>_{q}\eqno(9a)$$
$$\vert - (n_{1} -  1/2) >_{q}
\equiv {\frac {(a_{1q}^{\dagger})^{1-n_{1}} (a_{2q}^{\dagger})^{n_{1}}} 
 {([n_{1}]![1-n_{1}]!)^{1/2}}}\vert\tilde 0>_{q}\eqno(9b)$$
So the Hadamard transformation for q-deformed state is 
$$\vert n_{1} - 1/2>_{q} \rightarrow (-1)^{n_{1}} 
\vert n_{1} - 1/2>_{q} +\vert 1/2  - n_{1}>_{q}\eqno(10)$$
The usual Hadamard transformation for the  Jordan-Schwinger construction 
with usual oscillators is
$${\frac {(a_{1}^{\dagger})^{n_{1}} (a_{2}^{\dagger})^{n_{2}}} {{(n_{1}!n_{2}!)^{1/2}}}}\vert \phi>
\rightarrow (-1)^{(n_{1}-n_{2}+1)/2}{\frac {(a_{1}^{\dagger})^{n_{1}} (a_{2}^{\dagger})^{n_{2}}}
 {(n_{1}!n_{2}!)^{1/2}}}\vert \phi>$$
$$+{\frac {(a_{1}^{\dagger})^{n_{2}} (a_{2}^{\dagger})^{n_{1}}} {(n_{1}!n_{2}!)^{1/2}}}\vert \phi>
\eqno(11)$$
So the Hadamard transformation in terms of the q-deformed oscillators is:
$${\frac {(a_{1q}^{\dagger})^{n_{1}} (a_{2q}^{\dagger})^{n_{2}}} {{([n_{1}]![n_{2}]!)^{1/2}}}}
\vert\phi>_{q}\rightarrow (-1)^{(n_{1}-n_{2}+1)/2}                                     
{\frac {(a_{1q}^{\dagger})^{n_{1}} (a_{2q}^{\dagger})^{n_{2}}} {([n_{1}]![n_{2}]!)^{1/2}}}
 \vert\phi>_{q}$$
$$+{\frac {(a_{1q}^{\dagger})^{n_{2}} (a_{2q}^{\dagger})^{n_{1}}} {([n_{1}]![n_{2}]!)^{1/2}}}
\vert\phi>_{q}\eqno(12a)$$
Note that $n_{1},n_{2}$ is always $0$ or $1$ so as to correspond to the qubit. Hence the q-numbers 
$[n_{1}], [n_{2}]$ are always the usual numbers $n_{1}, n_{2}$ in our case. 
So $(12a)$ becomes
$${\frac {(a_{1q}^{\dagger})^{n_{1}} (a_{2q}^{\dagger})^{n_{2}}} {{(n_{1}!n_{2}!)^{1/2}}}}
\vert\phi>_{q}
\rightarrow (-1)^{(n_{1}-n_{2}+1)/2}                                     
{\frac {(a_{1q}^{\dagger})^{n_{1}} (a_{2q}^{\dagger})^{n_{2}}} {(n_{1}!n_{2}!)^{1/2}}}
\vert\phi>_{q}$$
$$+{\frac {(a_{1q}^{\dagger})^{n_{2}} (a_{2q}^{\dagger})^{n_{1}}} {(n_{1}!n_{2}!)^{1/2}}} 
\vert\phi>_{q}\eqno(12b)$$
Using $(1)$,$(7)$, and $n_{1}+n_{2}=1$ in $(12b)$ gives:
$$[F_{1}(\hat N_{1},q)a_{1}^{\dagger}]^{n_{1}} [F_{2}(\hat N_{2},q)a_{2}^{\dagger}]^{1-n_{1}}
\vert\phi>_{q}\rightarrow$$ 
$$(-1)^{n_{1}}[F_{1}(\hat N_{1},q)a_{1}^{\dagger}]^{n_{1}}
[F_{2}(\hat N_{2},q)a_{2}^{\dagger}]^{1-n_{1}}\vert\phi>_{q}$$
$$+ [F_{1}(\hat N_{1},q)a_{1}^{\dagger}]^{1-n_{1}} [F_{2}(N_{2},q)a_{2}^{\dagger}]^{n_{1}} 
\vert\phi>_{q}\eqno(13a)$$
where
$$F_{1}(\hat N_{1}, q)
=\sqrt{{\frac {q^{\hat N_{1}} \psi_{1} - q^{-\hat N_{1}}\psi_{2}} {{\hat N_{1} (q-q^{-1})}}}}\enskip,\enskip
F_{2}(\hat N_{2}, q)
=\sqrt{{\frac {q^{\hat N_{2}} \psi_{3} - q^{-\hat N_{2}}\psi_{4}} {{\hat N_{2} (q-q^{-1})}}}}
\eqno(13b)$$

For reasons already stated,the eigenvalues of the number operators are constrained to satisfy
$n_{1} + n_{2} = 1$ and the only possibilities are  $n_{1}=0, n_{2}=1$ or $n_{1}=1, n_{2}=0$.
The same restrictions also apply to usual (i.e.undeformed)  oscillators.
Hence we restrict the hatted number operators, 
$\hat N_{1}$ and $\hat N_{2}$,  by $\hat N_{1} + \hat N_{2} = I$ where $I$ is the identity operator.

In $(13b)$, $\psi_{i}(q)$ , $i=1,2,3,4$ are arbitrary functions of $q$ 
only  and $\psi_{i}(1)=1$. We take $\psi_{1}=\psi_{3}$ and $\psi_{2}=\psi_{4}$.
Also $\hat N_{1}+\hat N_{2}=I$. Under these circumstances we drop the suffixes from 
$F_{1}$ and $F_{2}$ and take the functional forms to be the same. 
This means that if one oscillator has the number operator as $\hat N$, 
the other oscillator  should be restricted to that  described by the number operator $I-\hat N$
($I$ ,the identity operator).The eigenvalues are $\hat n$ and $1-\hat n$ respectively ($\hat n=0,1$).
The harmonic oscillator realisations of the
q-oscillaters are described by the functions $F(\hat N,q)$and $F(1-\hat N,q)$.
Then $(13a)$ ,with $\hat n$ replacing $\hat n_{1}$ and
using $(1c)$,becomes
$$A\vert\eta > \rightarrow (-1)^{n}A\vert\eta> + B\vert\ -\eta>\eqno(14)$$
where 
$$\vert \eta> ={\frac {(a_{1}^{\dagger})^{n} (a_{2}^{\dagger})^{1-n}} {{(n!(1-n)!)^{1/2}}}}
\vert\phi>_{q};
~\vert -\eta> ={\frac {(a_{1}^{\dagger})^{1-n} (a_{2}^{\dagger})^{n}} {{(n!(1-n)!)^{1/2}}}}
\vert\phi>_{q} \eqno(15)$$
and $A=F(\hat N ,q)^{n}F(1+n - \hat N ,q)^{1-n}$ and
$B=F(1-\hat N ,q)^{1 -n}F(2-n- \hat N ,q)^{n}$.
For  $n=0$ this means 
$$F(\hat N,q) a_{2q}^{\dagger}\vert 0>_{1q}\vert 0>_{2q}$$
$$\rightarrow F(1-\hat N,q) a_{2q}^{\dagger}\vert 0>_{1q}\vert 0>_{2q}
+F(\hat N,q) a_{1q}^{\dagger}\vert 0>_{1q}\vert 0>_{2q}\eqno(16a)$$
\newpage
For $n=1$,
$$F(\hat N,q) a_{1q}^{\dagger}\vert 0>_{1q}\vert 0>_{2q}$$
$$\rightarrow  - F(\hat N,q) a_{1q}^{\dagger}\vert 0>_{1q}\vert 0>_{2q}
+F(1-\hat N,q) a_{2q}^{\dagger}\vert 0>_{1q}\vert 0>_{2q}\eqno(16b)$$
Obviously $(16a,b)$ would be indistinguishable from the usual Hadamard transformation
for $n=0,1$ if and only if    
$F^{-1}(\hat N, q) F(1-\hat N, q)= 1$. This operator equation 
written in terms of the eigenvalues $\hat n$ and $1-\hat n$ means 
$${\frac {\psi_{1}(q)}{\psi_{2}(q)}}={\frac {(q^{-\hat n}-\hat n q^{-\hat n}-\hat n q^{\hat n-1})}{(q^{\hat n}
-\hat n q^{\hat n}-\hat n q^{1-\hat n})}}
\eqno(17)$$
It is simple to check that $(17)$ is always true for $\hat n=0$ and $\hat n=1$ 
if $\psi_{1}(q)=\psi_{2}(q)=\psi(q)$ (say).
Therefore the Hadamard transformation can be realised with deformed qubits.

{\bf Case I}

There is only one arbitrary function $\psi(q)$ left and we now 
discuss its importance.First note that for $\psi_{1}=\psi_{2}=\psi_{3}=\psi_{4}=1$, $(2a,b)$
do not have any arbitrary parameter and just relates the opertors $a, a^{\dagger}$ with   
$a_{q}, a_{q}^{\dagger}$. Also from $(2b)$ we then have $N=\hat N$. This means that at the 
occupation number level the deformed states cannot be distinguished from the usual states.
So this is the realm of quantum computation with the usual "spin-up" and "spin-down" states. 

{\bf Case II}

But,$(2)$ is general if the arbitrary functions $\psi_{i}(q), i=1,2,3,4$ are
{\it not all equal to unity}. Then $N=\hat N - (1/s)~ln~ \psi(q)$ [$(2b)$]. 
Hence states labelled by the  occupation number are different as the eigenvalues of the
number operator of  usual oscillator states (i.e. usual quantum computation) and the
eigenvalues of the number operator of deformed oscillator states are now related by
$n=\hat n - (1/s)~ ln ~\psi(q)$. This would show up in the Jordan-Schwinger construction.

{\bf 4.Relation between the states in Case I with those in Case II}

Let us denote the angular momentum states in Case I by $\vert >_{I}$, and those in Case II by
$\vert >_{II}$.
Remembering that we have suppressed  $j=(n_{1}+n_{2})/2$ in the notation 
(since it is always $1/2$) and $m=(n_{1}-n_{2})/2=n_{1}-(1/2)$ and relabling
$n_{1}$ as $n$ etc. we have for Case I
$$\vert n-1/2>_{I}=\vert n>_{1}\vert 1- n>_{2}=\vert\hat n>_{1}\vert 1-\hat n>_{2}\eqno(18a)$$
or as $n=0,1$ and $n=\hat n$, the two states are 
$$\vert -1/2>_{I}=\vert \tilde 0>_{1}\vert \tilde 1>_{2} \enskip,\enskip \vert 1/2>_{I}=\vert \tilde 1>_{1}
\vert \tilde 0>_{2}\eqno(18b)$$
In Case II, the two states are 
$$\vert n'-1/2>_{II}=\vert n'>_{1}\vert 1-n'>_{2}=\vert\hat n - (1/s) ln \psi>_{1}
\vert 1-\hat n + (1/s) ln\psi>_{2}\eqno(19a)$$
$$\vert -(n'-1/2)>_{II}=\vert 1- n'>_{1}\vert n'>_{2}=\vert 1 -\hat n + (1/s) ln \psi>_{1}
\vert \hat n - (1/s) ln\psi>_{2}\eqno(19b)$$

However, here $n'=\hat n -(1/s) ln \psi(q)$, and  the two states are
$$\vert -1/2>_{II}=\vert \tilde 0>_{1}\vert \tilde 1>_{2}
=\vert\hat n - (1/s) ln \psi>_{1}\vert 1-\hat n+(1/s) ln\psi>_{2}\eqno(19c)$$
$$\vert 1/2>_{II}=\vert \tilde 1>_{1}\vert \tilde 0>_{2}=\vert 1 -\hat n + (1/s)ln \psi>_{1}
\vert \hat n - (1/s)ln\psi>_{2}\eqno(19d)$$
{\it Consistency} now demands that 
$$\hat n = (1/s)ln \psi(q)\eqno(20)$$
(Note that the state on the left-hand side of the equations $(18),(19)$
are angular momentum states, while the right-hand sides are the direct product
of oscillator states.)

{\bf 5.An alternate formalism for quantum computation}

First consider Case I,i.e. $(18)$.
It is immediately evident that so far as quantum computation is concerned
nothing much is gained by choosing these states because the
eigenvalue of the  number operators for usual and deformed oscillators are
identical. {\it So it will be impossible to distinguish the states in Case I
from those of usual oscillators at the level of experimental
realisations or consequences}.

Now consider Case II, i.e. Eq.$(19)$. 

(a)We have (for $n'=\hat n - (1/s)ln \psi(q)$)
$$\vert n'>_{II}=\psi(q)^{(n_{1}+n_{2})/2} 
 {\frac {(a_{1q}^{\dagger})^{n_{1}} (a_{2q}^{\dagger})^{n_{2}}} 
 {([n_{1}]![n_{2}]!)^{1/2}}} \vert \tilde 0>_{q}
 =\psi(q)^{(n_{1}+n_{2})/2}\vert n>_{I}=\psi (q)^{1/2}\vert n>_{I}\eqno(21)$$
Therefore,
$$ \frac {_{II}<n'\vert n'>_{II}} {{_I}<n\vert n>_{I}}=\psi (q)\eqno(22)$$ 
{\it This means that the states in Case II can be distinguished from those in 
Case I or from the usual oscillator states at the level of experimental 
realisations or consequences.} 

(b)$\hat n= (1/s) ln \psi(q)$ means $\psi (q)= =e^{s\hat n}=q^{\hat n}$,
$\hat n$ is the
eigenvalue of the number operator and hence $\hat n \geq 0$ while $0<s<1$.
Here $\hat n$ cannot be zero because then we will have $\psi(q)=1$ i.e.
Case I. So here $\hat n >0$.
This means that {\it the deformed states in Case II can be related to any
usual oscillator states with occupation 
number greater than zero.}This is a very rich theoretical structure and opens up
enormous possibilities for experimental realisations and consequences by suitably 
choosing the two parameters $\hat n$ and  $s$.

{\bf 6. The Phase Shift transformation}

Let us now consider the Phase Shift transformation of qubit states defined as usual:
$\vert x >\rightarrow e^{i x \theta}\vert x >$ which in our notation is
$\vert n-{\frac 12}>\rightarrow e^{in\theta}\vert n - {\frac 12}>$
where $\theta$ is the phase shift.So denoting initial and final states by $i,f$ 
$$\vert n-1/2>_{If}=e^{in\theta}\vert n-1/2>_{Ii}\eqno(23a)$$
$$\vert n-1/2>_{IIf}=e^{in\theta}\vert n-1/2>_{IIi}
=e^{in\theta}q^{\hat n/2}\vert n-1/2>_{Ii}=q^{\hat n/2}\vert n-1/2>_{If}\eqno(23b)$$
Then for $n=0$, $\vert -{\frac 12}>_{I,II}\rightarrow \vert -{\frac 12}>_{I,II}$ 
and for $n=1$, $\vert {\frac 12}>_{I,II}\rightarrow e^{i\theta}\vert {\frac 12}>_{I,II}$.
Hence the phase shift transformation can also be implemented for a single deformed qubit.
Moreover, note that the two cases I and II can be distinguished from the fact that 
$$\frac {_{IIf}<n-1/2\vert n-1/2>_{IIf}} {{_If}<n-1/2\vert n-1/2>_{If}}=\psi (q)=q^{\hat n}=e^{s\hat n}\eqno(24)$$ 
So here also the presence of the function $\psi(q)=q^{\hat n}=e^{s\hat n}$ gives two parameters 
(a)a positive integer $\hat n>0$ and (b)a positive fraction $s$ where $0<s<1$ that can be exploited
for both experimental realisations and consequences.

{\bf 7.Conclusion}

Thus, we have shown that so far as realisation of the single qubit
Hadamard and Phase Shift gates are concerned, {\it q-deformed qubit states 
can also be used}. A principal advantage over the usual formalism is the 
occurrence of an arbitrary function of the deformation parameter $q=e^{s}$.
This function is $\psi(q)= q^{\hat n}= e^{s\hat n}$. So we have two free parameters
(i)$s$,  $0<s<1$    and  (ii) $\hat n > 0$ . These 
can be used to determine whether observed experimental realisations
of theoretical predictions obtained from the usual formalism are fully
satisfactory  or not.If not, then these parameters can be exploited to see whether 
corrections to the results can be calculated.These aspects require further 
investigations, but the very possibility that quantum computation may also be 
done using {\it q-deformed qubits} is indeed appealing.
Whether the difference between quantum computation using usual spin states 
and quantum computation using q-deformed qubit states  
is susceptible to experimental observations in the NMR realisation of quantum
logic gates [18] is an interesting problem in its own right.

\end{document}